\documentclass[12pt]{iopart}

\usepackage{graphics}
\usepackage{graphicx}
\usepackage{epsfig}
\begin{document}

\title[IPTA Challenge 1]{Applying Bayesian Inference to the first International Pulsar Timing Array data challenge}

\author{Neil J. Cornish}

\address{Department of Physics, Montana State University, Bozeman,
MT 59717, USA}

\begin{abstract}
This is a very brief summary of the techniques I used to analyze the IPTA challenge 1 data sets. I tried many things, and
more failed than succeeded, but in the end I found two approaches that appear to work based on tests
done using the open data sets. One approach works directly with the time domain data, and the other
works with a specially constructed Fourier transform of the data. The raw data was run through TEMPO2
to produce reduced timing residuals for the analysis. Standard Markov Chain Monte Carlo
techniques were used to produce samples from the posterior distribution function for the model parameters.
The model parameters include the gravitational wave amplitude and spectral slope, and the white noise
amplitude for each pulsar in the array. While red timing noise was only included in Dataset 3, I found
that it was necessary to include effective red noise in all the analyses to account for some of the
spurious effects introduced by the TEMPO2 timing fit. This added an additional amplitude and slope parameter
for each pulsar, so my overall model for the 36 pulsars residuals has 110 parameters. As an alternative to
using an effective red noise model, I also tried to simultaneously re-fit the timing model model while
looking for the gravitational wave signal, but for reasons that are not yet clear, this approach was
not very successful. I comment briefly on ways in which the algorithms could be improved. My best
estimates for the gravitational wave amplitudes in the three closed (blind) data sets are:
(1) $A=(7.3\pm 1.0)\times 10^{-15}$; (2) $A=(5.7\pm 0.6)\times 10^{-14}$; and (3) $A=(4.6\pm 1.3)\times 10^{-15}$.
\end{abstract}



\maketitle

\section{Method}

A description of the IPTA data challenge and the data sets can be found at {\it http://www.ipta4gw.org/}. The
first round consisted of three parts - Datasets 1,2, and 3, each with one open data set for testing purposes
and one closed data set as a blind challenge. The datasets were arranged in increasing order of difficulty, starting
with evenly sampled data with bright gravitational wave signals and moving on to unevenly sampled data with
weak gravitational wave signals. The IPTA data challenges are similar to the
Mock LISA Data Challenges~\cite{Arnaud:2006gm}, the main difference being that the people that put this challenge
together have real detectors to work with as well.

My approach to the analysis was to use Bayesian inference and Markov Chain Monte Carlo sampling
to produce posterior distributions for the
gravitational wave amplitude and the pulsar timing noise. The Bayesian approach requires a noise model -
which fixes the form of the likelihood - a signal model, and priors on the model parameters. For stationary,
Gaussian noise the likelihood is given by
\begin{equation}\label{like}
p(d\vert s, n) = \frac{1}{\sqrt{(2\pi)^L {\rm det}C}}\exp\left(-\frac{1}{2} \sum_{(\alpha i)(\beta j)} r_{(\alpha i)}
C^{-1}_{(\alpha i)(\beta j)} r_{(\beta j)}\right) \, ,
\end{equation}
where $C$ is the covariance matrix, which depends on the noise in the detectors (in this case the individual
Earth-pulsars arms) and the amplitude of the stochastic gravitation wave background, and $r = d-s$ denotes the residuals
after a signal model $s$ has been subtracted from the data $d$. Here Greek indices $\alpha={1,N}$ label which
detector (Earth-pulsar arm), and Latin indices $i={1,M}$ label the data samples. This expression for the likelihood
applies to any collection of gravitational wave detectors with stationary, Gaussian noise, and is widely used in
theoretical studies of ground-based and space-based gravitational wave
detectors~\cite{Flanagan:1993ix, Finn:1997qx, vanHaasteren:2008yh, Adams:2010vc}.
It applies to any data representation, including time domain, frequency domain and wavelet domain. For evenly
sampled time domain data with colored noise it is best to transform to the frequency domain where
$C_{(\alpha i)(\beta j)} = C'_{\alpha\beta}(i) \delta_{ij}$, as this leads to substantial savings when
computing the likelihood: $\sim M N^3$ versus $\sim M^3 N^3$ floating point operations.

In the current setting we are looking for a stochastic gravitational wave signal, so we don't have a deterministic signal
model for the GW component of the data. Instead, we are looking for correlations in the signals between Earth-pulsar pairs
$\alpha,\beta$, which depend on the angle $\theta_{\alpha\beta}$ between the line of sight to each pulsar and follow the
Hellings-Downs~\cite{Hellings:1983fr}
correlation curve $H_{\alpha\beta}$, which has the form
\begin{equation}
H_{\alpha\beta} = \frac{1}{2}+\frac{3(1-\cos\theta_{\alpha\beta})}{4} \ln\left(\frac{1-\cos\theta_{\alpha\beta}}{2}\right)-\frac{1-\cos\theta_{\alpha\beta}}{8}+\frac{1}{2}\delta_{\alpha\beta} \, .
\end{equation}
We do however have a deterministic timing model that accounts for effects such as the pulsar spin-down, sky location and proper
motion, and a host of other effects, such as relativistic time delays due to the Earth's motion about the Sun. The leading terms
in the timing model have the form
\begin{equation}\label{time}
\fl s(t) = a_0 + a_1 t + a_2 t^2 + a_3 \cos(\omega_0 t+\phi_0)+ a_4 t \cos(\omega_0 t+\phi_1)+ a_5 \cos(2 \omega_0 t+\phi_2) +\dots
\end{equation}
The parameters $a_0,a_1,a_2$ are used to fit the pulsar spin-down, $a_3$ and $\phi_0$ account for the sky location and leading
order time dilation effects, which introduce periodic variations with $\omega_0 = 2\pi/{\rm year}$, $a_4$ and $\phi_1$
account for pulsar proper motion, while $a_5$ and $\phi_2$ account for second order time dilation effects and the eccentricity of
the Earth's orbit. The publicly available TEMPO2~\cite{Hobbs:2006cd, Edwards:2006zg, Hobbs:2009yn} software package
fits a more involved timing model to each individual
pulsar using a chi-squared goodness of fit. Taking the TEMPO2 residuals and subtracting the raw data that was
provided for the open data sets reveals a clear excess of power with a spectral slope of $\sim -4$, in addition to a
wide bump around $f=1/{\rm year}$ from the $a_3$ and $a_4$ terms, along with a narrow spike at $f=2/{\rm year}$ from
the $a_5$ term.

A naive approach would be to take the TEMPO2 timing residuals and compare the
the cross correlation between pulsar pairs to the Hellings-Downs correlation curve. The problem with this approach is
that the TEMPO2 fitting procedure introduces spurious features in the residuals, potentially removing or at least masking
any gravitational wave signal that might be present. A better approach is to combine the fitting of the timing model with the
cross-correlation analysis. I tried this approach, but the results were not entirely satisfactory. For some reason the recovered
GW amplitude always tended to get biased high. I suspect that I might be missing some important contributions to the
timing model. I plan to re-visit this approach in an effort to understand why it didn't work as expected.

Simply ignoring the features introduced by TEMPO2 and using a model that only accounts for the GW signal and white noise
in Datasets 1 and 2, or the red and white noise of Dataset 3, leads to GW amplitudes that appear to be biased high (based on
analysis of the open data sets). The one caveat here is that I am not 100\% sure about wether the quoted amplitudes for the
GW signal refer to the astrophysical signal or the effective level registered in the detector. The astrophysical level $S_h$ is
related to the observed level in an Earth-Pulsar link $S^{\rm obs}_h$ by $S^{\rm obs}_h = \frac{2}{3} S_h$ (the two-thirds
factor comes from sky and polarization averaging). This corresponds to an amplitude ratio bias of $\sqrt{2/3}=0.816$. If the
factor of two-thirds was not included, then my results for the GW amplitude need to be multiplied by a factor of 1.225.
In any case, this factor is not large enough to account for the bias that I saw with the naive analysis, which showed a
$\sim 40\%$ amplitude bias. To check my analysis codes in a more controlled fashion, I wrote my own simulation software,
using the same approach for generating stochastic backgrounds that we used in the Mock LISA Data Challenges~\cite{Babak:2008aa}
(based on a adding together the response to independent stochastic signals from different patches of the sky using a
fine Healpix sky grid). This simulation confirmed that the two-thirds factor should be included, and
using my own simulated data, with no TEMPO2 processing, I was able to perfectly recover the
injected GW amplitude and spectral slope, and the pulsar noise levels. 

Since the TEMPO2 fitting is done on a per-pulsars basis, the errors in the timing models should be uncorrelated between
pulsars. I therefore decided to model the timing model errors as an additional, uncorrelated red noise contribution
to each pulsar, described by a spectral amplitude $S_r$ and slope $\gamma_r$. For Dataset 1 we were told that all
the pulsars shared the same white noise level, so for that analysis the model I used had $2 N +3=75$ parameters (since there
are $N=36$ pulsars in the simulated array). For Dataset 2 the white noise level in each pulsar was different, so the model
had $3N+2=110$ parameters. For Dataset 3 there is an additional red noise component, but I assumed that this was already
taken care of by the red noise model that I introduced to account for the TEMPO2 fitting noise, so again the model had 110
parameters. Moreover, if I understood the statement about the red noise level that was injected, it is essentially irrelevant to
the analysis, as it lies below the fiducial 100 ns white timing noise for frequencies above 10 nHz (Here I am using the fact that
100 ns white timing noise corresponds to a spectral density of $S_n= 2.417e-8\;{\rm Hz}^{-3}$, and the red noise was
quoted at $S_r = 5.77e-22 f^{-1.7}\;{\rm Hz}^{-3}$, which cross at $f=(5.77e-22/2.417e-8)^{0.59}= 9.7 {\rm nHz}$, which is
at the very edge of the frequency band for a 5 year data set, and in a region where the fit to the pulsar spin-down removes
any meaningful signal). In most runs I fixed the gravitational wave spectral slope at the expected value of
$\gamma=-11/3$ to improve the amplitude recovery. 

With the model established, the next step is to decide on how to compute the likelihood. The simplest and most
robust approach is to implement (\ref{like}) in the time domain, following the approach
of van Haarsteren {\it et al.}~\cite{vanHaasteren:2008yh}. The downside of this approach is that the matrix
inversion is very computationally expensive for
an $L$ by $L$ matrix with $L= N \times M = 36 \times 130 = 4680$ rows and columns. Since
the correlation matrices are positive definite, they
can be inverted by way of Cholesky factorization. The matrix determinant is given by product of the squares of the diagonal
elements of the Cholesky decomposition. I used the {\em LAPACK} routines dpotrf and dpotri to perform the factorization and
inversion. The codes were run on multi-core Mac Pro workstations, where an interesting side-effect of having the
{\em LAPACK} routines as built in functions is that the matrix inversion is automatically parallelized, taking up an average
of 4 cores. Even with the parallel processing, each likelihood evaluation took $\sim 6$ seconds, making it difficult to
produce long Markov Chains. To adequately explore the posterior distribution we need many samples in the chains, with the
required number of samples scaling roughly linearly with the model dimension. The combination of the large model dimension and
the high cost of generating the chains leads to long runs times (or the need to secure larger computing resources!). To
shorten the ``burn-in'' phase of the analysis (where the chains locate the region of high posterior mass), I first ran on
each pulsar individually using a three parameter model consisting of the white noise level, the gravitational wave amplitude (the
slope was fixed at $\gamma=-13/3$), and the low frequency cut-off in the spectrum. The spectral estimates from
the individual pulsar runs were used as a starting point for the full run on the pulsar network.
The low frequency cut-off parameter requires some explanation. The time domain auto-correlation function for time
lags $\tau$, $C(\tau)$, forms a Fourier pair with the frequency domain
spectrum $S(f)$:
\begin{equation}
C(\tau) = \int_{f_{\rm low}}^\infty S(f) e^{i 2\pi f \tau} df \, .
\end{equation}
Setting $f_{\rm low} = 0$ for power-law power spectra leads to improper correlation functions. As
van Haarsteren {\it et al.}~\cite{vanHaasteren:2008yh} point out,
the correlation function should not depend very strongly on $f_{\rm low}$ so long as $f_{\rm low} < 1/T_{\rm obs}$, as the fitting
of the pulsar spin-down model removes the part of the spectrum that causes the divergence in the limit $f_{\rm low} \rightarrow 0$.
However, I found that the value of $f_{\rm low}$ did affect the analysis quite significantly, so I made it a parameter to be
determined from the data. The chains generally favored values around $f_{\rm low} = 0.7/T_{\rm obs}$. The correlation
functions used in the full analysis have the form
\begin{equation}
C^w_{(\alpha i)(\beta j)} = (\delta t_\alpha)^2\, \delta_{\alpha\beta}\, \delta_{ij} \, ,
\end{equation}
for white noise with amplitude $\delta t_\alpha$, and
\begin{eqnarray}\label{gw}
C^h_{(\alpha i)(\beta j)} &=& -H_{\alpha\beta}\frac{2 A^2 f_{\rm low}^{\gamma+1}}{3(2\pi)^2 {\rm year}^{\gamma+3}}\left(
\Gamma(\gamma+1)\cos(\pi(\gamma+3)/2)(2\pi f_{\rm low}\tau)^{-(\gamma+1)} \right. \nonumber \\
&&\left. +\sum_{n=0}^Q (-1)^n \frac{(2\pi f_{\rm low}\tau)^{2n}}{(2n)\! (2n+\gamma+1)} \right) \, ,
\end{eqnarray}
for a GW signal with amplitude $A$ at $f=1/{\rm year}$ and spectral slope $\gamma$. Here
$\tau=\vert t_{(\alpha i)}-t_{(\beta j)} \vert$, and the sum over $n$ needs to extend to $Q=12$ if
$f_{\rm low}$ is allowed to go as high as $1/T_{\rm obs}$. The injected GW signals have spectral slope $\gamma=-13/3$.
An expression similar to (\ref{gw}) applies to the un-correlated red noise, but with $H_{\alpha\beta}$ replaced
by $\delta_{\alpha\beta}$.

In the runs where I included a timing model of the form (\ref{time}) for each pulsar, the MCMC alternated between updating the
parameters in the correlation matrix (which costs of order $\sim M^3 N^3$ floating point operations), and updating the
parameters in the timing model (which costs of order $\sim M^2 N^2$ floating point operations). Because the timing model
updates are so much cheaper, it made sense to do many timing updates for each correlation matrix update. So while the
timing model involves many more parameters than the spectral model, we are able to spend many more iterations exploring the
much larger timing model parameter space for little additional computational cost. However, as I remarked earlier, the
runs that used the timing model did not do as good a job of recovering the amplitude of the GW signal as did the runs with
an uncorrelated red noise model.
 
In an attempt to speed up the calculation of the likelihood, I also considered frequency domain implementations of the likelihood
calculation (\ref{like}). Transforming the data to the frequency domain turned out to be a highly non-trivial exercise
due to the very red spectra and the
uneven data sampling. Either one of these issues is easy to overcome on its own, but in tandem they are a bitch. The approach
I settled on combined several techniques that have been used previously in pulsar timing data analysis. To handle the unevenly
sampled data I used an tempered MCMC algorithm to fit the timing residuals to a sum of sines and cosines using the log likelihood
\begin{equation}\label{c2}
\log L = -\left(\sum_{i=1}^M\left[d_i - \sum_{j=1}^{P}( a_j \cos(j\omega t_i)+b_j \sin(j\omega t_i))\right]\right)^2/\delta t^2
\, , 
\end{equation}
with $\delta t = 1$ ns (the choice of $\delta t$ is largely irrelevant since I used a tempered MCMC). To avoid the Gibbs phenomenon
from the finite-time window function and the very red spectra, I set $\omega = 2\pi/T$ with $T > T_{\rm obs}$. I found that
$T = 1.1 T_{\rm obs}$ worked well, and that somewhat larger and smaller values also worked. The sum over frequencies has to
be extended beyond the usual Nyquist frequency because of the un-even time-sampling. Using the Fourier coefficients $a_j$ and
$b_j$ directly gave reasonable results, but I found that it was better to re-map the data onto the true time interval with
uniform time sampling, then transform the data with a standard FFT. To avoid the Gibbs phenomena I applied a zero
phase shift double difference filter in the time domain: $d'[i] =d[i-1]+d[i+1]-d[i]$, and compenstated by dividing the
Fourier amplitudes derived from the $d'$ sequence by the transfer function $2(\cos(2\pi f \Delta t)-1)$. Applying this
filtering procedure to the full data set leads to aliasing of high frequency power
to low frequencies, so I used the fit derived from (\ref{c2}) to first split the data into high-passed and low-passed
components. The low frequency part was transformed with the filter, while the high frequency part was transformed without
the filter. The results were then added together to give the complete Fourier transform. The sum over frequencies in (\ref{like})
was restricted to the range $[2/T_{\rm obs},50/T_{\rm obs}]$. The lowest frequency bin was dropped as the fitting of the pulsar
spin-down model takes out most of the very low frequency power, and including the $1/T_{\rm obs}$ frequency bin was found to
bias the GW amplitude low. I dropped the highest frequencies in part because they contain no information about the GW amplitude,
and also because the interpolation of the data becomes less trustworthy one short time scales.
For unevenly sampled data, the Fourier components are not strictly orthogonal, and $C_{(\alpha i)(\beta j)} \neq
C'_{\alpha\beta}(i) \delta_{ij}$. To keep the analysis simple I ignored this complication and assumed orthogonality.

For the priors I used uniform distributions in the log of the spectral amplitudes, and uniform distributions in the spectral
slope parameters. I used a very wide prior range, and found that the distributions stayed well within these ranges, with
the exception of the time domain analysis of the closed Dataset3 data, where the low end of the GW prior was seen to cut-off
the distribution, which suggests that the GW level was not confidently detected in that analysis.

I found that in most of the challenge data sets the GW amplitude was sufficiently high that there was no question about
wether a detection had been made. The one exception was the closed data from Dataset3. The GW amplitude flirted with zero
during the time and frequency domain runs, indicating that a pure noise model may have greater Bayesian evidence.
To test this I ran a trans-dimensional Reversible Jump Markov Chain Monte Carlo (RJMCMC) algorithm on the frequency domain
data that allowed for transitions between a GW + noise model and a pure noise model. The ratio of the time that the chains spent in each model provides the Bayes factor, or evidence
ratio for the two models. If our prior belief is that the two models are equally likely, then this ratio is the betting odds
for one model versus the other. As we shall see, the odds turned out to favor the GW model by a fairly comfortable margin
(but perhaps by not enough to send a paper to Nature if this was real data).

\section{Results}

\begin{figure}[t]
\vspace*{0.5in}
\begin{center}
\epsfig{file=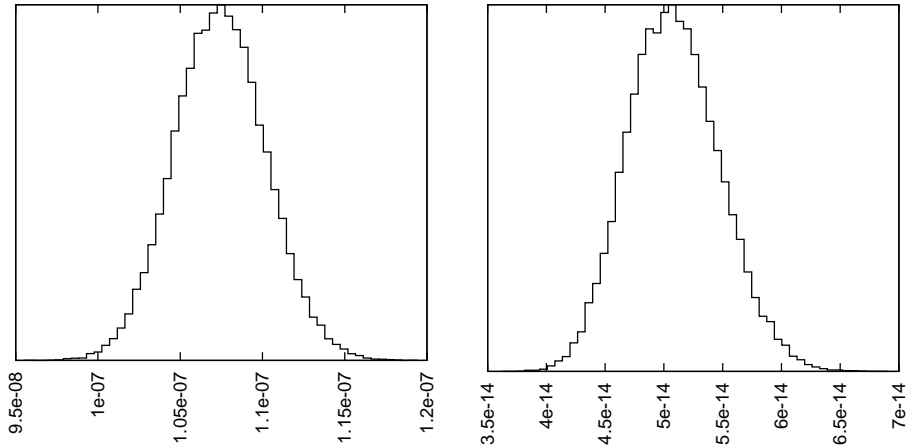,  width=0.8\linewidth}
\end{center}
\vspace*{0.1in}
\caption{Posterior distributions for the level of the white timing noise (left panel) and the
dimensionless gravitational wave amplitude (right panel) for the open Dataset1 data.}
\label{fig:open1}
\end{figure}

I begin by showing results from the frequency domain analysis. The GW spectral slope was held fixed at $\gamma = -13/3$
for the analyses shown in Figures 1-7. Figure 1 shows the posterior distributions for the
level of the white timing noise (in seconds) and the dimensionless gravitational wave amplitude $A$ using the
open data from Dataset 1. The white noise level came out a little above the injected value of $10^{-7}$ s,
while the gravitational wave amplitude peaked very close to the injected value of $A=5\times 10^{-14}$.
Figure 2 shows the same quantities, but now for the closed Dataset 1 data. This time the white noise level was
a fraction below 100 ns. The GW amplitude distribution peaked at $A=7.3\times 10^{-15}$, which is significantly
lower than the level seen in the open data set. 

\begin{figure}[t]
\vspace*{0.5in}
\begin{center}
\epsfig{file=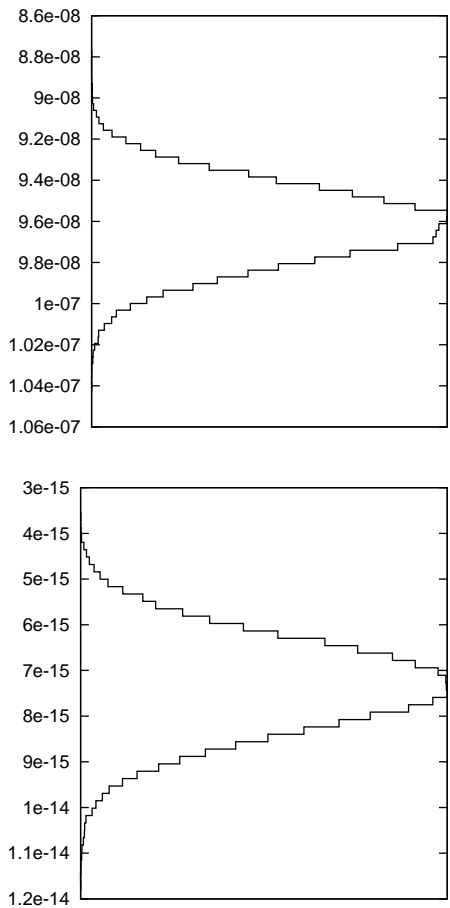, width=0.8\linewidth}
\end{center}
\vspace*{0.1in}
\caption{Posterior distributions for the level of the white timing noise (left panel) and the
dimensionless gravitational wave amplitude (right panel) for the closed Dataset1 data.}
\label{fig:closed1}
\end{figure}

Figure 3 shows the posterior distributions for the dimensionless gravitational wave amplitude for Dataset2 for both the
open and closed data. The posterior distribution peaks very near the injected value of $A=5\times 10^{-14}$ for the open
data set, and the distribution for the closed data set peaks a little higher at $A=5.7\times 10^{-14}$.

\begin{figure}[t]
\vspace*{0.6in}
\begin{center}
\begin{tabular}{cc}
\epsfig{file=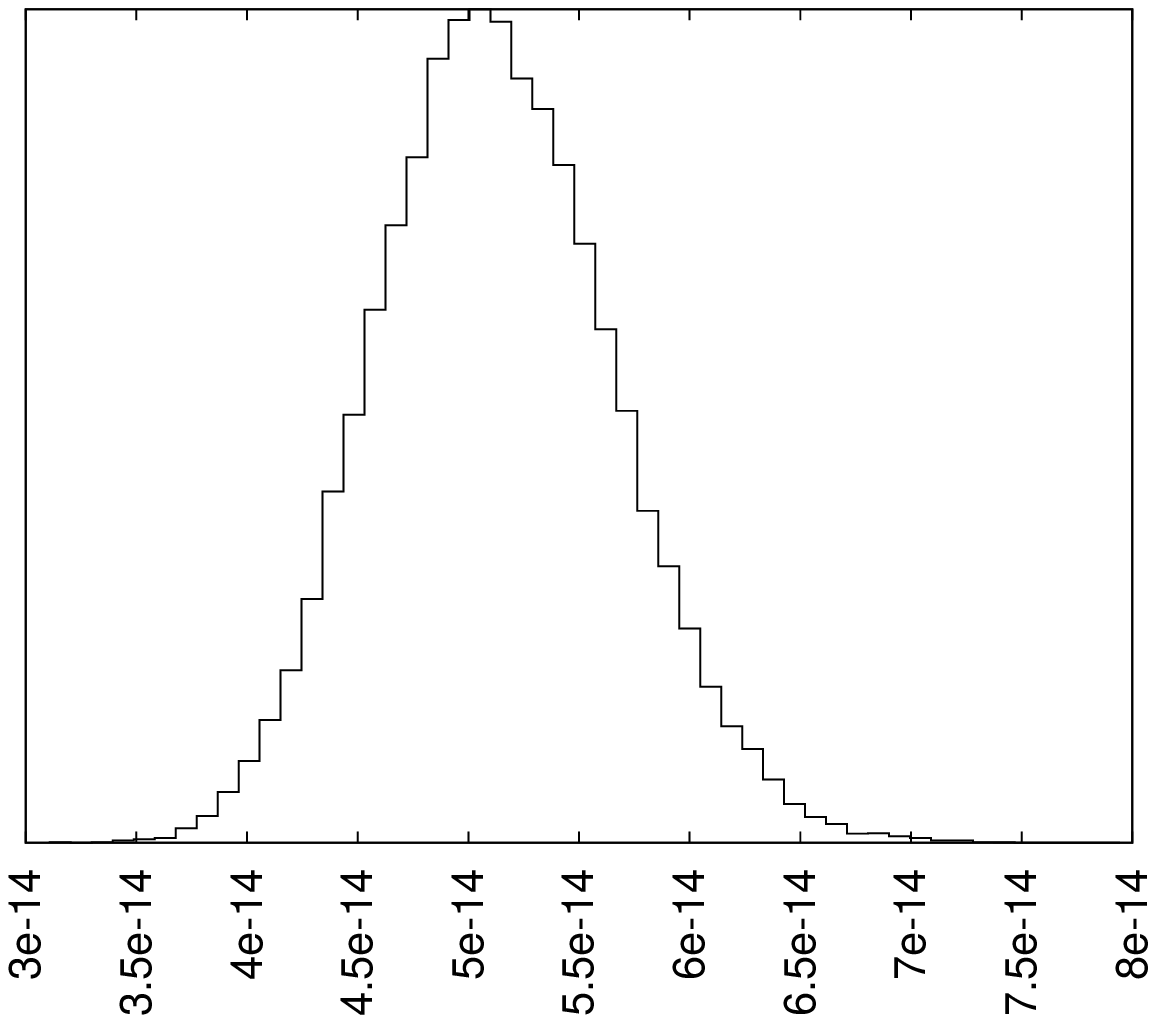,width=0.4\linewidth} & 
\epsfig{file=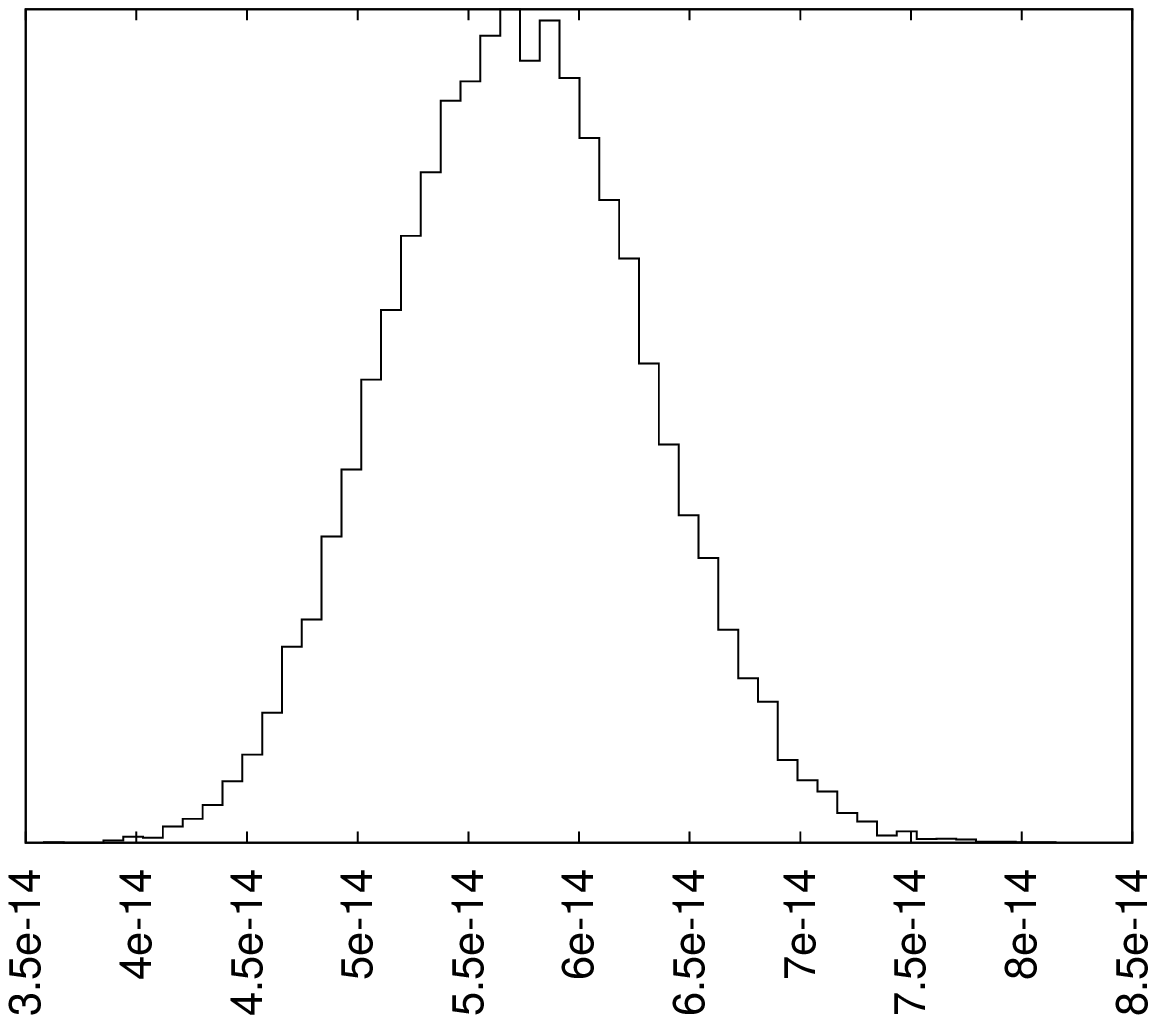,width=0.4\linewidth} 
\end{tabular}
\end{center}
\vspace*{0.1in}
\caption{Posterior distributions for the gravitational wave amplitude for the Dataset2 open (left)
and closed (right) data.}
\label{fig:data2}
\end{figure}

Figure 4 shows the posterior distributions for the dimensionless gravitational wave amplitude for Dataset3 for
both the open and closed data. The posterior distribution peaks very near the injected value of
$A=1\times 10^{-14}$ for the open data set, and the distribution for the closed data set peaks
quite a bit lower at $A=4.6\times 10^{-15}$. The standard deviation for the closed data set is
$\sigma_A=1.3\times 10^{-15}$, so in frequentist terms we have a ``3.5 sigma'' detection. Using a
RJMCMC algorithm I found that the evidence ratio in favor of a gravitational wave signal being present
was $(180 \pm 20):1$, or a little less than ``3 sigma'' in frequentist terms.

\begin{figure}[t]
\vspace*{0.5in}
\begin{center}
\begin{tabular}{cc}
\epsfig{file=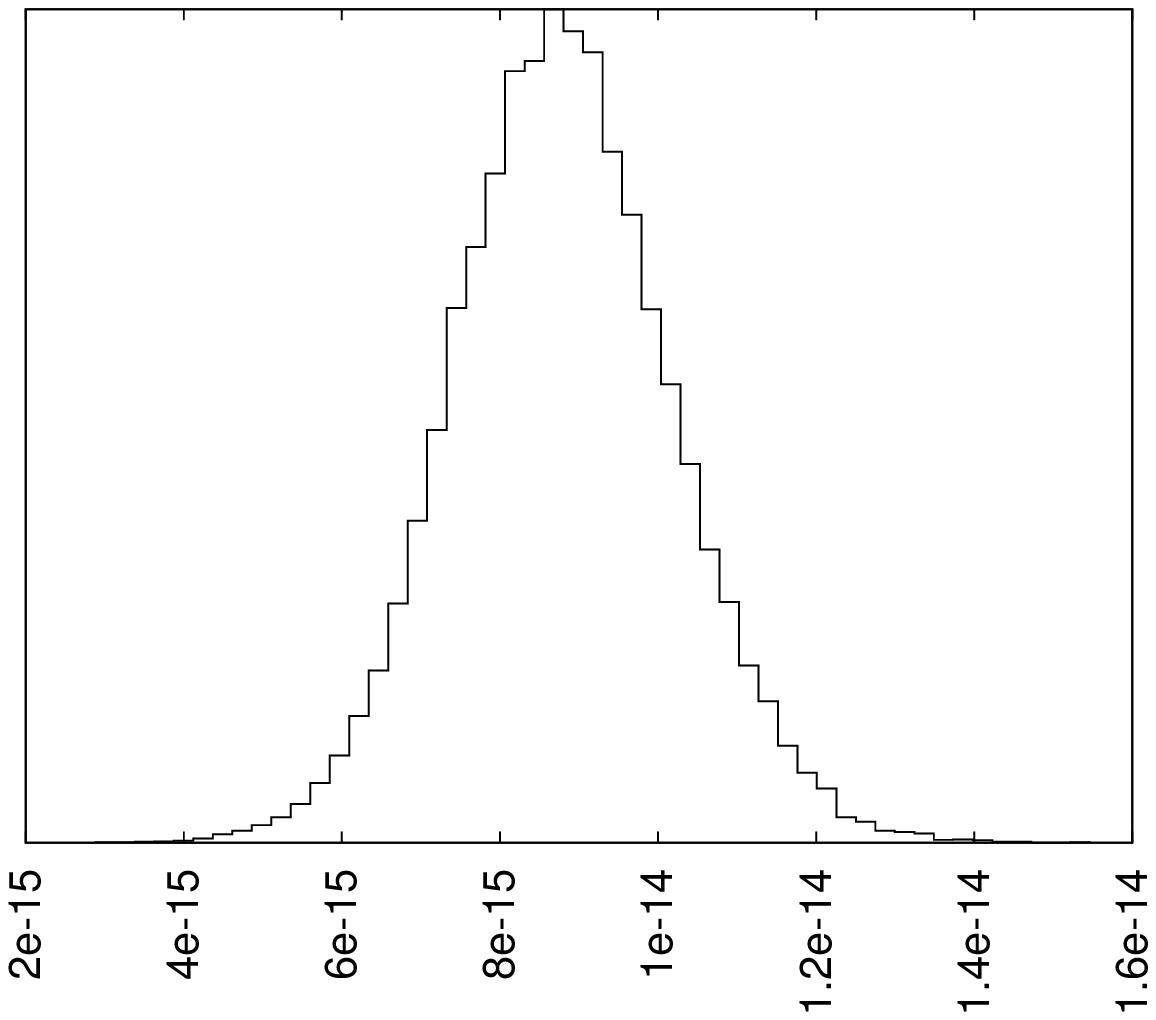,width=0.4\linewidth} & 
\epsfig{file=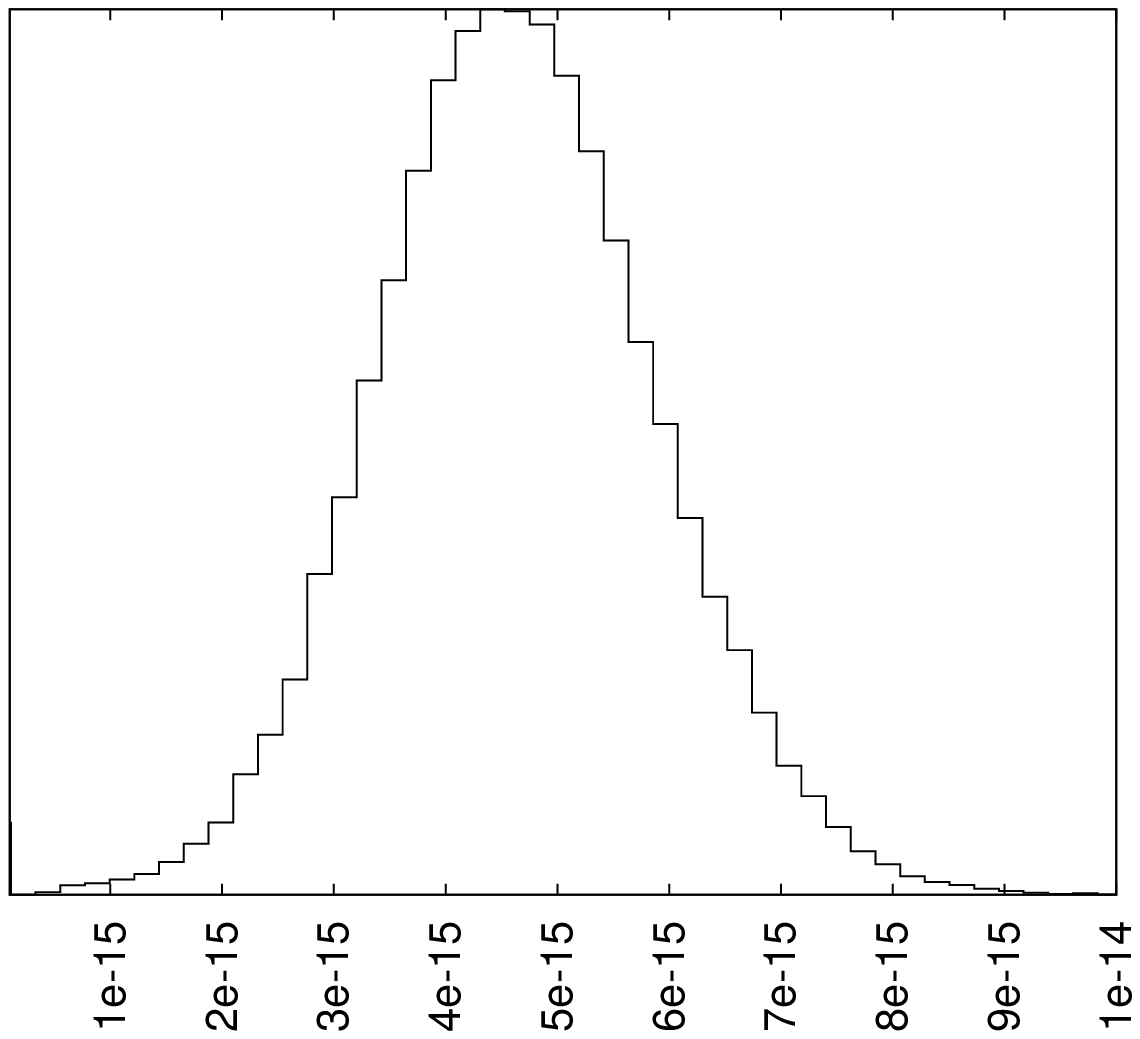,width=0.4\linewidth} 
\end{tabular}
\end{center}
\vspace*{0.1in}
\caption{Posterior distributions for the gravitational wave amplitude for the Dataset3 open (left)
and closed (right) data.}
\label{fig:data3}
\end{figure}

The time domain analysis appears to have a bias towards low gravitational wave amplitudes. This can probably
be attributed to the pulsar spin-down fit removing a large portion of the low frequency power, and this
removal not being accounted for in the model for the GW power spectrum. The same bias was easily avoided in
the frequency domain analysis by simply discarding the very low frequency data. A possible remedy in the time
domain would be to modify the model for the GW power spectrum such that the spectrum turns over at low frequencies.
I suspect that the $f_{\rm low}$ parameter in the current model attempts to achieve a similar result. Unfortunately,
the time domain analyses take days or weeks to run, so it isn't easy to try out a lot of alternative strategies.
I present the current results from the time domain codes, but with the caveat that I am fairly certain that the
levels are biased low by $\sim 20\%$ or more.

Figure 5 shows the posterior distributions for the dimensionless gravitational wave amplitude for Dataset1 for the
open and closed data using a time domain analysis. The posterior distribution for the open data set 
peaks at $A=4.1\times 10^{-14}$, which is quite a bit below the injected value. The distribution for the closed
data set peaks at $A=5.9\times 10^{-15}$, which is below the $A=7.3\times 10^{-15}$ value found by
the frequency domain analysis.

\begin{figure}[t]
\vspace*{0.5in}
\begin{center}
\begin{tabular}{cc}
\epsfig{file=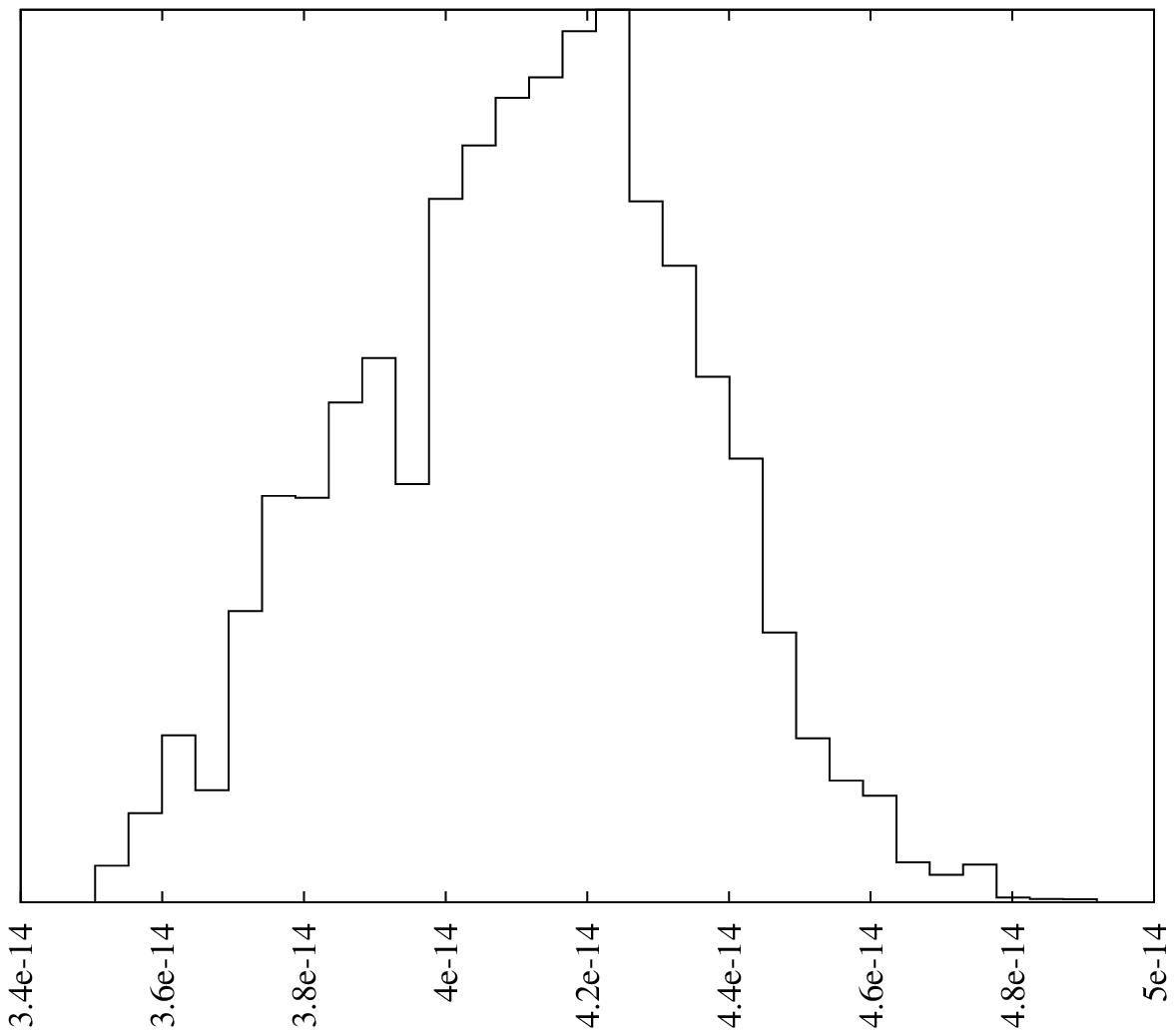,width=0.4\linewidth} & 
\epsfig{file=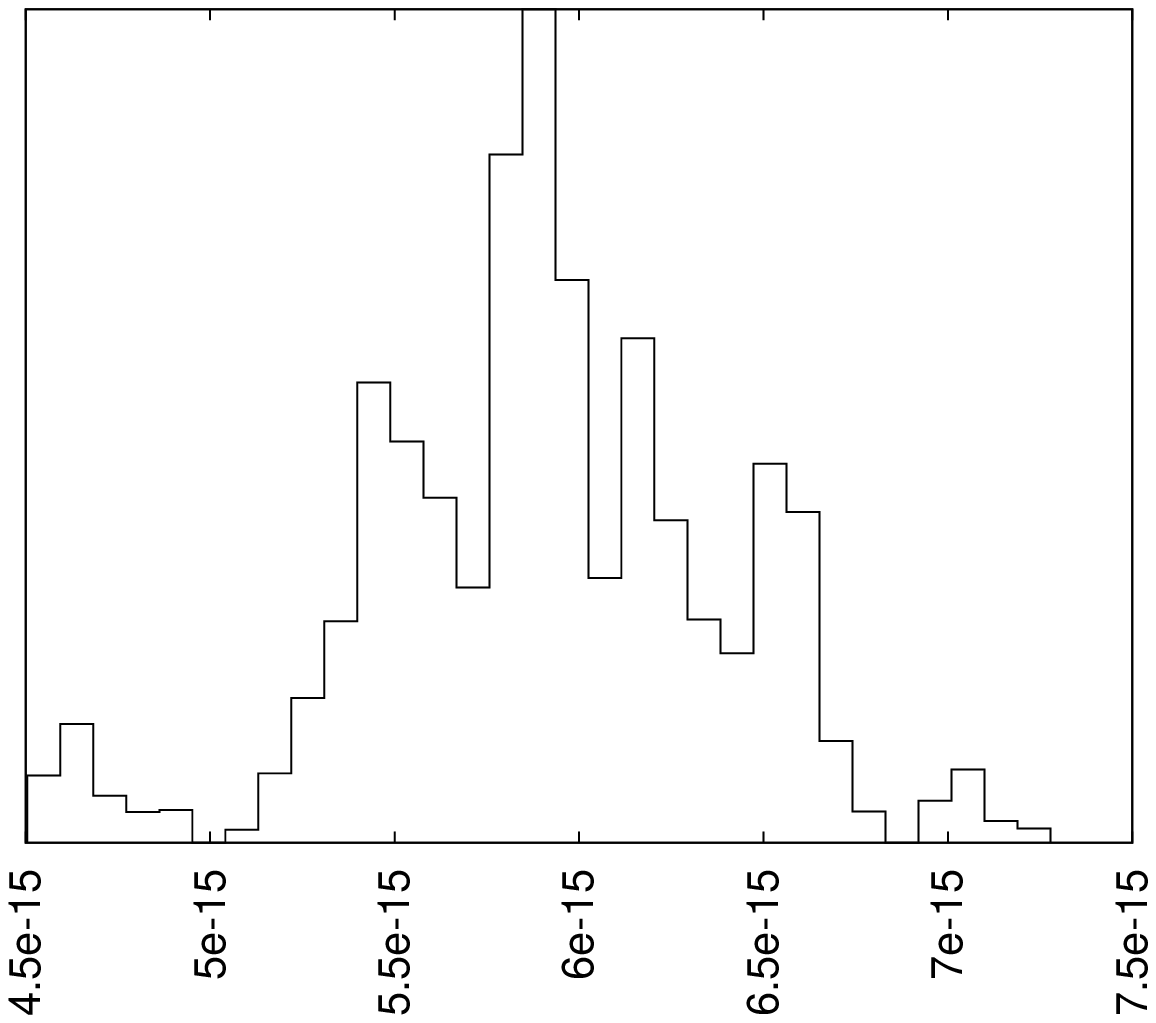,width=0.4\linewidth} 
\end{tabular}
\end{center}
\vspace*{0.1in}
\caption{Posterior distributions for the gravitational wave amplitude from a time domain analysis of Dataset1 for the
open (left) and closed (right) data.}
\label{fig:data1t}
\end{figure}

Figure 6 shows the posterior distributions for the dimensionless gravitational wave amplitude for Dataset2 for the
open and closed data using a time domain analysis. The posterior distribution for the open data set 
peaks at $A=4.6\times 10^{-14}$, which is below the injected value. The distribution for the closed
data set peaks at $A=4.7\times 10^{-14}$, which is below the $A=5.7\times 10^{-14}$ value found by
the frequency domain analysis.

\begin{figure}[t]
\vspace*{0.5in}
\begin{center}
\begin{tabular}{cc}
\epsfig{file=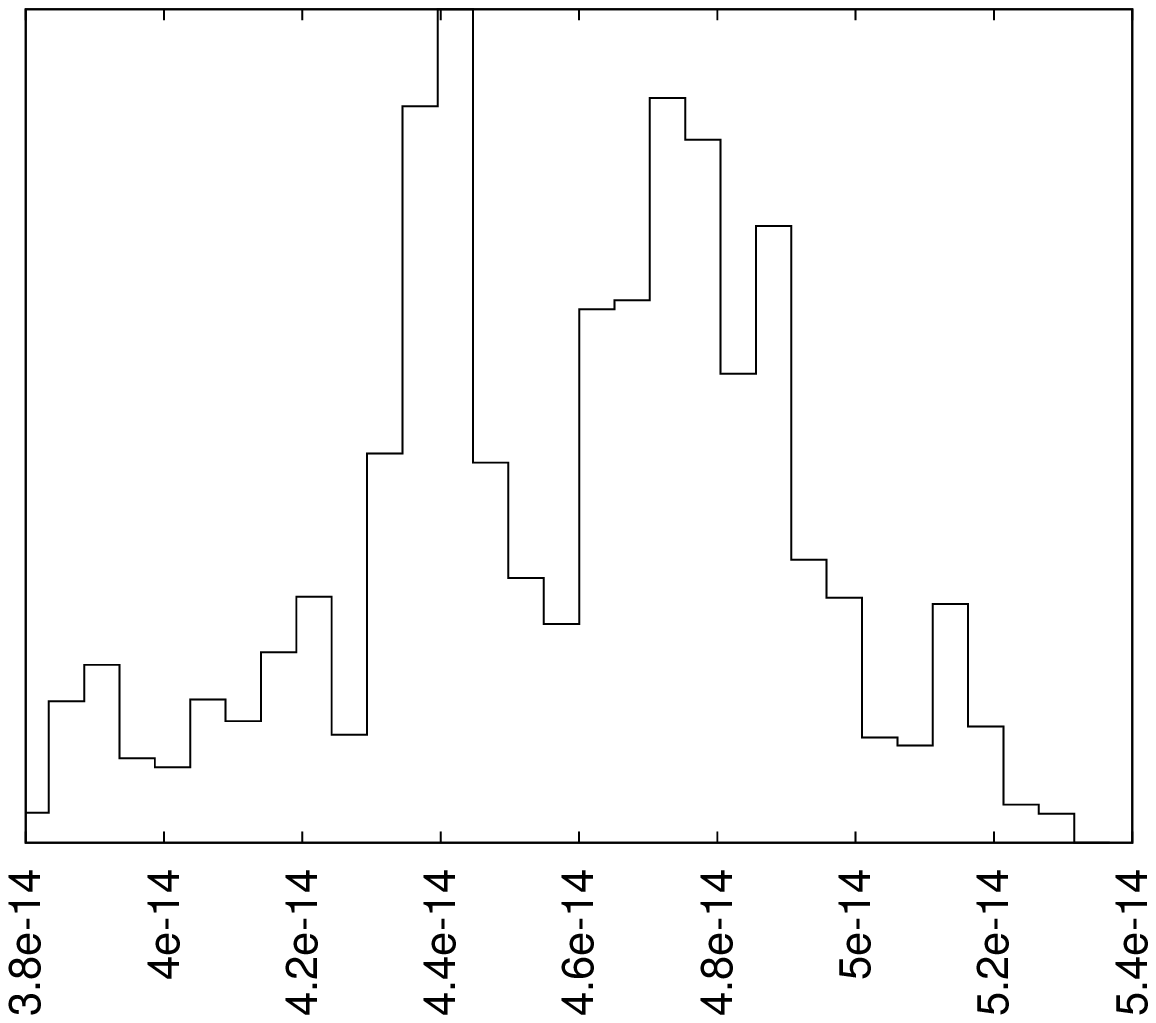,width=0.4\linewidth} & 
\epsfig{file=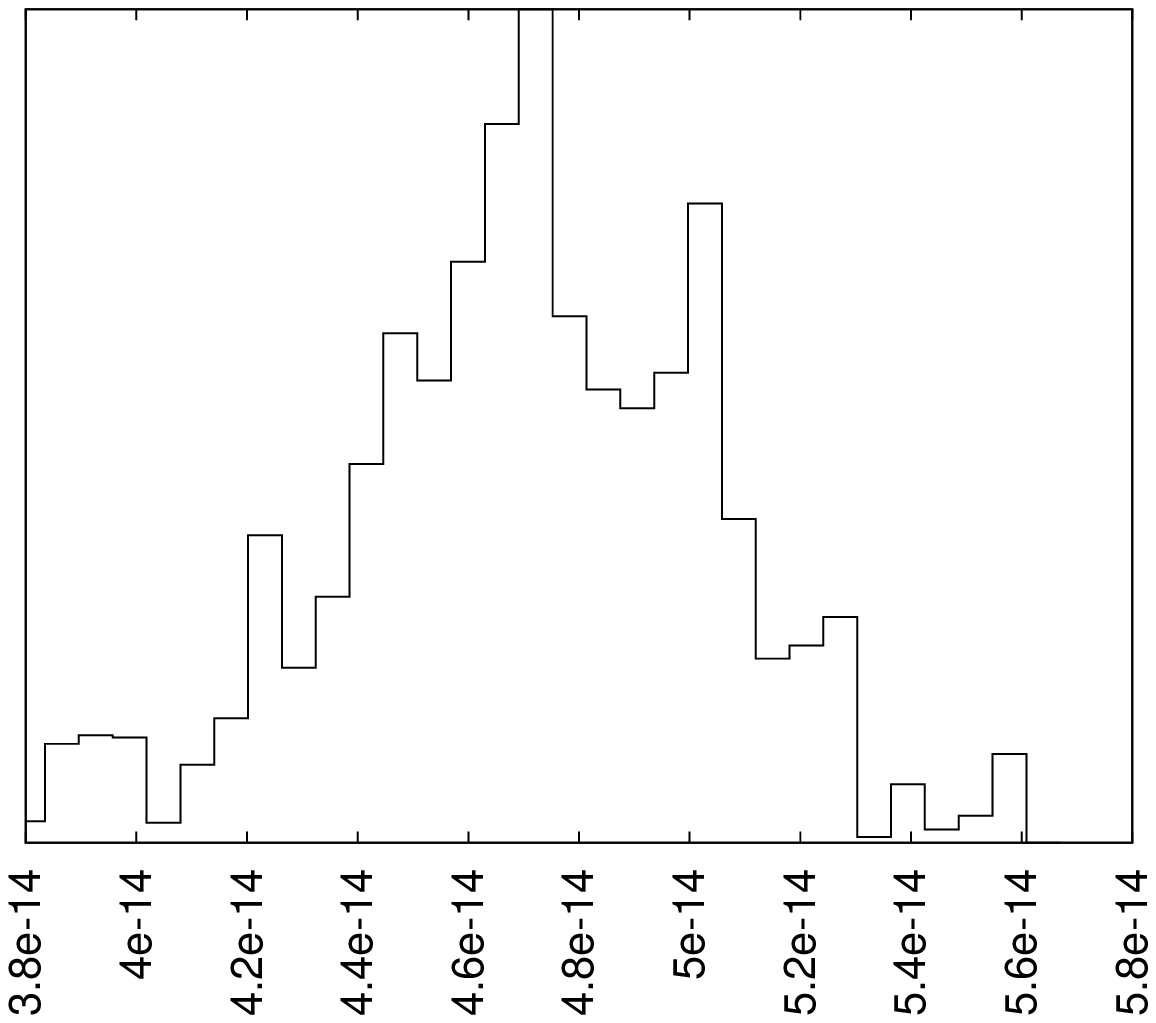,width=0.4\linewidth} 
\end{tabular}
\end{center}
\vspace*{0.1in}
\caption{Posterior distributions for the gravitational wave amplitude from a time domain analysis of Dataset2 for the
open (left) and closed (right) data.}
\label{fig:data2t}
\end{figure}

Figure 7 shows the posterior distributions for the dimensionless gravitational wave amplitude for Dataset3 for the
open and closed data using a time domain analysis. The posterior distribution for the open data set 
peaks at $A=5.5\times 10^{-15}$, which is significantly below the injected value. The distribution for the closed
data set runs up against the boundary of the prior range, and it is not clear that detection has been made in this
case.

\begin{figure}[t]
\vspace*{0.5in}
\begin{center}
\begin{tabular}{cc}
\epsfig{file=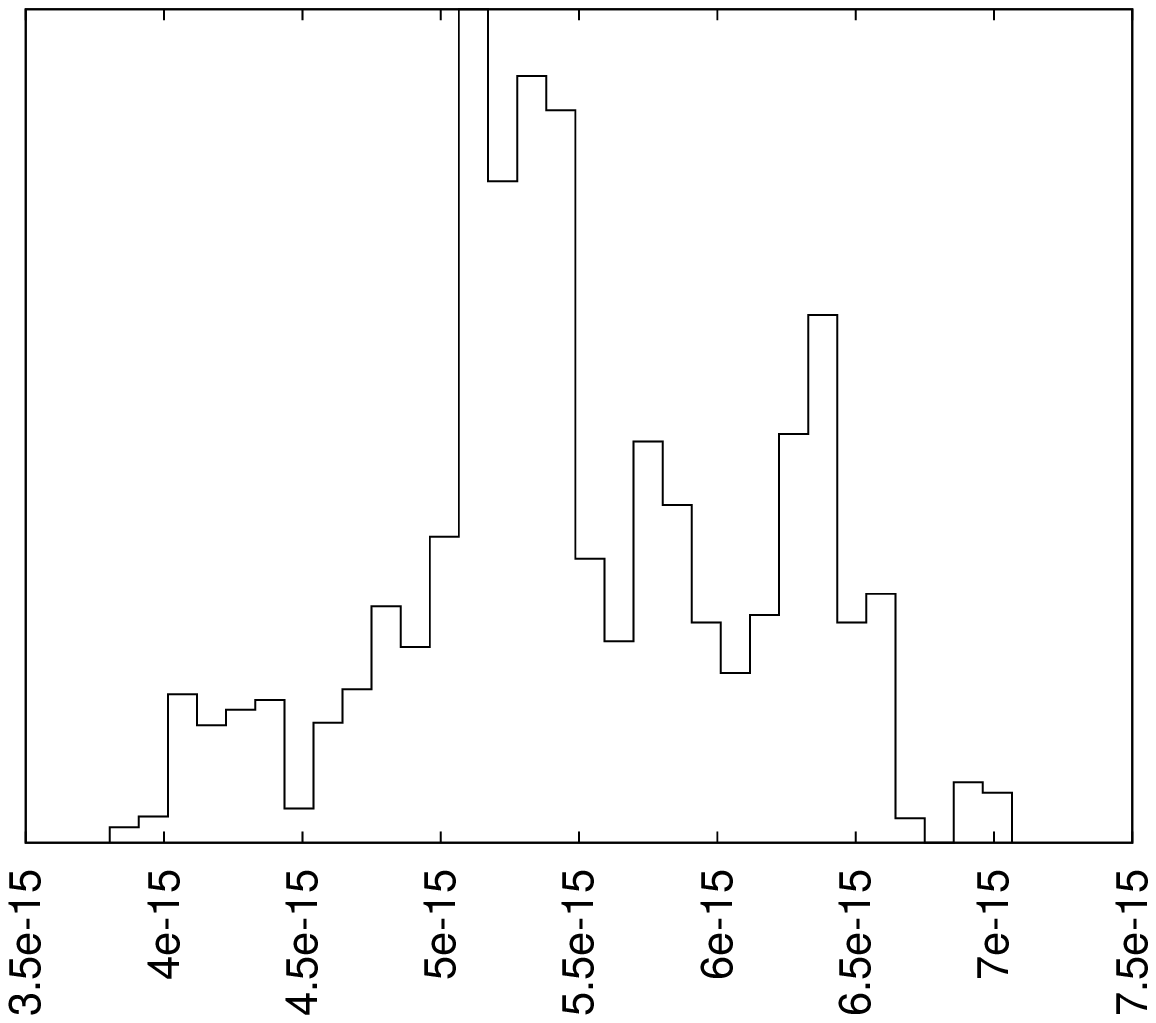,width=0.4\linewidth} & 
\epsfig{file=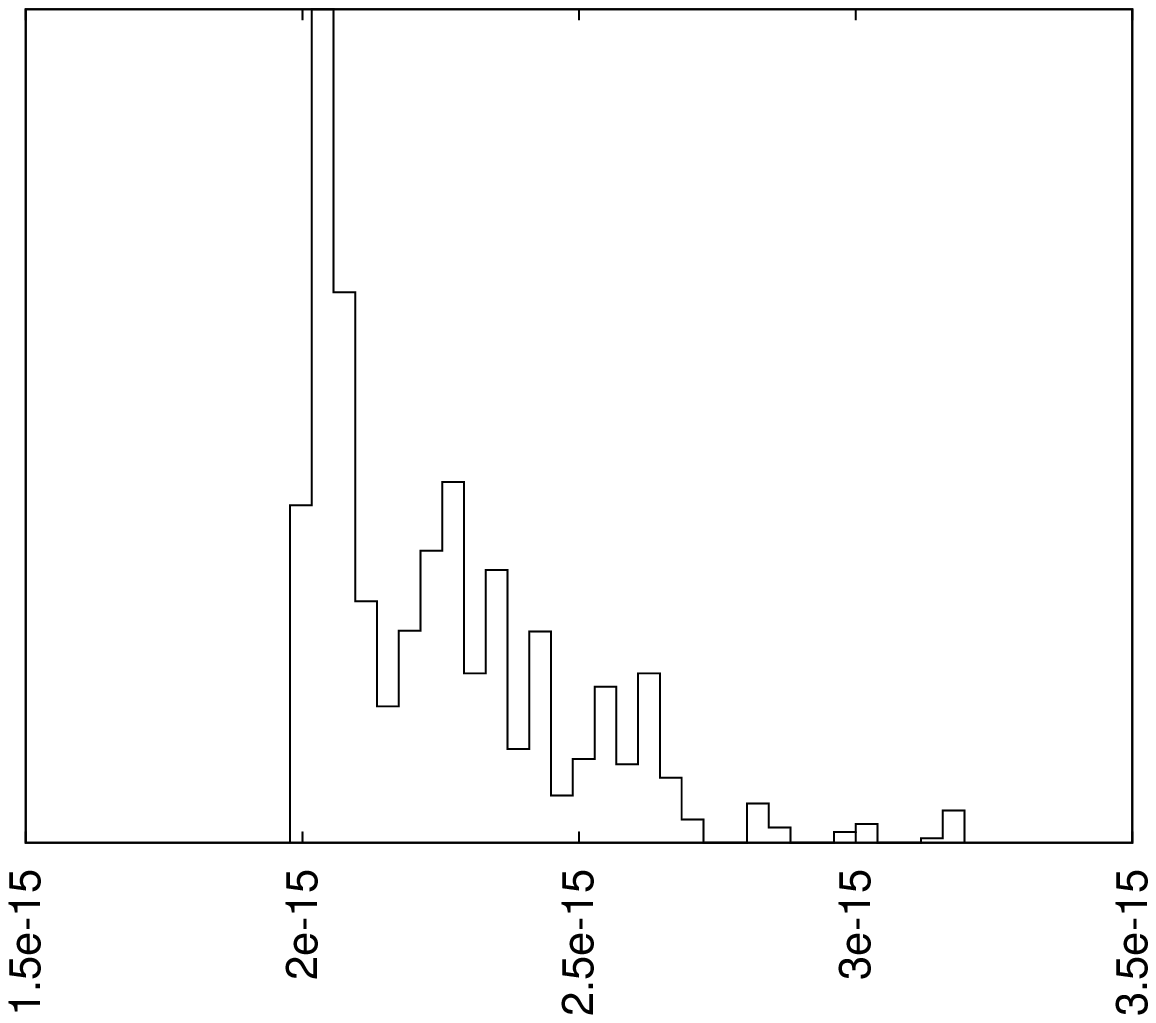,width=0.4\linewidth} 
\end{tabular}
\end{center}
\vspace*{0.1in}
\caption{Posterior distributions for the gravitational wave amplitude from a time domain analysis of Dataset3 for the
open (left) and closed (right) data.}
\label{fig:data3t}
\end{figure}

Returning to the frequency domain analysis, and now letting the spectral slope of the GW spectrum vary, I found the
posterior distributions for the spectral slope $\gamma$ and the GW amplitude $A$ shown in Figure 8. As expected, the
spread in the amplitude distribution is wider than when $\gamma$ was held fixed. The distribution for $\gamma$ is biased
a little high relative to the injected value of $\gamma=-13/3$, which may be due to the procedure used to perform the
Fourier transform, or more likely, to interplay with the red noise model used to account for the timing model errors.

\begin{figure}[t]
\vspace*{0.5in}
\begin{center}
\epsfig{file=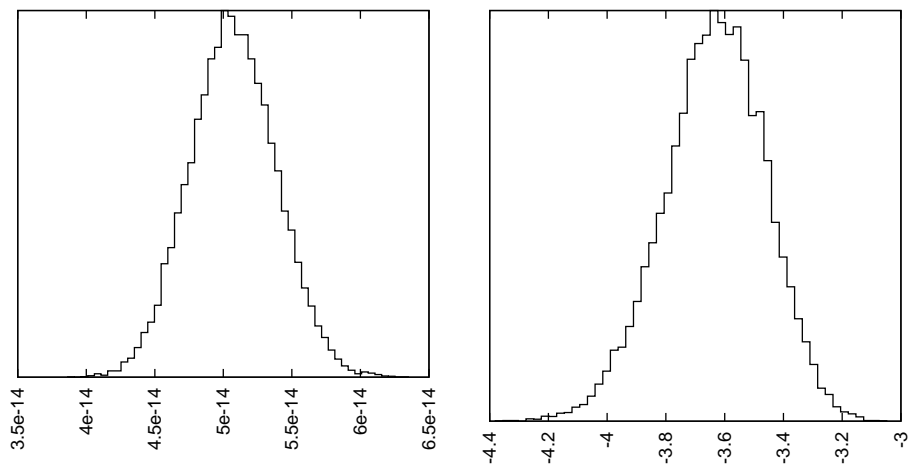,  width=0.8\linewidth}
\end{center}
\vspace*{0.1in}
\caption{Posterior distributions for the gravitational wave  amplitude (left panel) and the  spectral slope (right panel) for the open Dataset1 data.}
\label{fig:open1s}
\end{figure}

\section{Discussion}

I was able to get reasonable results from the time domain and frequency domain analyses. There does appear to be a bias
in the time domain estimates for the gravitational wave amplitude, which I attribute to a mis-match between the spectral
model (a simple power law) and what is likely present in the data (a power law that flatens out at low frequencies). This
should be easy to fix with a suitable change to the spectral model.

I remain puzzled as to why a simultaneous fit to the timing model and the gravitational wave correlation function did not
produce better results. This could be due to an implementation error, or it may be that the this approach is flawed in
some more serious way. The frequency domain analysis suggests that modeling the errors in the timing model as
red ``modeling'' noise is quite effective. I am still a little uneasy about the frequency domain analysis as the
procedure I used to transform the data was fairly Byzantine.

As a general rule I don't like to specify the model used in the analysis in too rigid a fashion. I prefer to let the
data select the best model. In future I would like to use a full RJMCMC analysis that is able to transition between
models. For example, when I look at the distributions for some of the red noise parameters it is clear that they are
poorly constrained by the data. A RJMCMC analysis would automatically discard these parameters from the model. This
approach may also cure some of the ills of the timing model analysis, where again it was clear that many of the timing
model parameters were poorly constrained.

I have some ideas about how to speed up the interminably slow time domain analysis. There the bottle neck comes from
having to invert a large non-sparse matrix. Suppose however that we have found an approximate form for the matrix.
We can then use the Cholesky factors to transform to a data representation where the a samples are approximately
uncorrelated in time and space ({\it i.e.} in both $\alpha,\beta$ and $i,j$). In this transformed representation
``nearby'' correlation matrices will be almost diagonal, and fast perturbative methods can be used to
compute their inverse.

\section{Acknowledgments}
This work was supported by NASA grant NNX07AJ61G.

\end{document}